\documentclass[twocolumn,showpacs]{revtex4}
\usepackage{times,xspace}
\usepackage{amsbsy,amssymb,amsmath,bm}
\usepackage{graphicx,color,epsfig,rotate}
\usepackage{fancyhdr}

\def\bbbc{{\mathchoice {\setbox0=\hbox{$\displaystyle\rm C$}\hbox{\hbox
to0pt{\kern0.4\wd0\vrule height0.9\ht0\hss}\box0}}
{\setbox0=\hbox{$\textstyle\rm C$}\hbox{\hbox
to0pt{\kern0.4\wd0\vrule height0.9\ht0\hss}\box0}}
{\setbox0=\hbox{$\scriptstyle\rm C$}\hbox{\hbox
to0pt{\kern0.4\wd0\vrule height0.9\ht0\hss}\box0}}
{\setbox0=\hbox{$\scriptscriptstyle\rm C$}\hbox{\hbox
to0pt{\kern0.4\wd0\vrule height0.9\ht0\hss}\box0}}}}

\newcommand{\ignore}[1]{}
\newcommand{\mComment}[1]{}
\newcommand{\gComment}[1]{}
\newcommand{\jComment}[1]{}
\newcommand{\rComment}[1]{}
\newcommand{\lComment}[1]{}

\renewcommand{\mComment}[1]{\textcolor{blue}{Manny: #1}}
\renewcommand{\gComment}[1]{\textcolor{red}{Gerardo: #1}}
\renewcommand{\jComment}[1]{\textcolor{green}{Jim: #1}}
\renewcommand{\rComment}[1]{\textcolor{magenta}{Ray: #1}}
\renewcommand{\lComment}[1]{\textcolor{purple}{Rolando: #1}}

\begin{document}

\title{A Derivation of the Fradkin-Shenker Result From Duality:
Links to Spin Systems in External Magnetic Fields and Percolation 
Crossovers}

\author{Zohar Nussinov}
\email{zohar@viking.lanl.gov}
\affiliation{Theoretical Division,
Los Alamos National Laboratory, Los Alamos, NM 87545}

\date{Received \today }

\begin{abstract}
In this article, we illustrate how the qualitative phase diagram
of a gauge theory coupled to matter can be directly proved and how
rigorous numerical bounds may be established. 
Our work reaffirms the seminal result of 
Fradkin and Shenker from another vista. Our main ingredient is  
the combined use of the self-duality of
the three dimensional Z2/Z2 theory and 
an extended Lee-Yang theorem. We comment
on extensions of these ideas and firmly
establish the existence of a sharp crossover 
line in the two dimensional Z2/Z2 theory.
\end{abstract}
\maketitle

\section{Introduction}

Lattice gauge theories \cite{Kogut} 
witnessed an accelerated revival 
in condensed matter physics during the last decade. Their applications
are widespread. Amongst others, these include 
novel theories of liquid crystals \cite{LRT},
the incorporation of Berry phase effects in quantum spin 
systems \cite{subir}, \cite{qcp}, and stimulating 
suggestions for long-distance physics of 
lightly doped Mott-Hubbard insulators \cite{SF}. Further research
relating to fundamental questions in gauge theories
followed, e.g. \cite{KNS}, \cite{Nagaosa}. 
Central to many of these investigations 
is the behavior of matter fields minimally coupled to 
gauge fields. Several key results in 
these theories were noted long ago by Fradkin and 
Shenker \cite{Fradkin}
(complemented by treatments in \cite{banks}). Perhaps the 
best known result  of \cite{Fradkin} is the demonstration that  
(when matter fields carry the fundamental unit of charge) 
the Higgs and confinement phases of gauge theories are 
smoothly connected to each other and are as different 
as a liquid is from a gas. This result remains one of 
the cornerstones of our understanding of the phases 
of gauge theories. Although derived 
long ago, the physical origin of this effect 
does not seem to be universally 
agreed upon. 

In the current article, we revisit this old 
result and rederive it for the original
${\mathbb Z}_{2}/{\mathbb Z}_{2}$ theory investigated in \cite{Fradkin}.
Our proof relies merely on duality
and the Lee-Yang theorem. We further 
illustrate why similar results are 
anticipated for other gauge 
theories. Our derivation highlights
the origin of this phenomenon as
akin to the absence of phase transitions
in spin systems in a magnetic
field. Notwithstanding the 
absence of true non-analyticities,
some such spin models display
a percolation crossover line \cite{kertesz}
at which the surface tension 
of an oppositely oriented spin cluster
vanishes. In this article, we firmly establish the existence of precisely 
such {\bf{a sharp percolation crossover}} line for one of the most 
trivial ${\mathbb Z}_{2}/{\mathbb Z}_{2}$ theories 
(the d = 1+1 dimensional theory).

\section{${\mathbb Z}_{2}$ Matter Coupled to ${\mathbb Z}_{2}$ 
Gauge Fields}
\label{Z2/Z2}

In matter coupled gauge theories, matter fields $(\{\sigma_{i}\}$)
reside as lattice sites $i$ while gauge fields $U_{ij}$ reside
on the links connecting sites $i$ and $j$.
The ${\mathbb Z}_{2}$ matter coupled to ${\mathbb Z}_{2}$ gauge field 
theory (${\mathbb Z}_{2}/{\mathbb Z}_{2}$ in common notation) 
is the simplest incarnation of a matter coupled gauge theory. 
Its
action reads
\begin{eqnarray}
S = - \beta \sum_{\langle i j \rangle} \sigma_{i} U_{ij} \sigma_{j} 
- K \sum_{\Box} UUUU 
\label{intro}
\end{eqnarray} 
on a hypercubic lattice. Here, 
the first sum is over all nearest neighbor links $\langle i j \rangle$ 
in the lattice while the second is the product
of the four gauge fields $U_{ij} U_{jk} U_{kl} U_{ki}$
over each minimal plaquette (square) of the lattice.
Both matter ($\sigma_{i}$) and gauge ($U_{ij}$) 
fields are Ising variables within the ${\mathbb Z}_{2}/{\mathbb Z}_{2}$ 
theory: $\sigma_{i} = \pm 1$, $U_{ij} = \pm 1$. 
A trivial yet fundamental observation
is that the quantity $z_{ij}
 \equiv \sigma_{i} U_{ij} \sigma_{j}$, where $i$ and $j$ denote
two nearest neighboring lattice sites, is invariant
under local ${\mathbb Z}_{2}$ gauge transformations
\begin{eqnarray}
\sigma_{i} \to \eta_{i} \sigma_{i}, ~  U_{ij} \to \eta_{i} U_{ij} \eta_{j}
\end{eqnarray}
with the arbitrary on-site $\eta_{i} = \pm 1$ \cite{Josephson}.
The action of Eq.(\ref{intro})
may be trivially written in terms of these 
gauge invariant bond variables
$\{z_{ij}\}$ as
\begin{eqnarray}
S = - \beta \sum_{links} z_{ij} - K \sum_{\Box} zzzz. 
\label{z}
\end{eqnarray}
The matter coupling $\beta$ acts as a magnetic 
field on the spin variable $z$.
On a new lattice whose sites reside on the centers of 
all bonds, this is none other
than a model having 4-spin 
interactions augmented
by a ${\mathbb Z}_{2}$ symmetry breaking
(for finite $\beta$) magnetic field.
For $\beta >0$, the link expectation value 
$\langle z_{ij} \rangle \equiv
\langle \sigma_{i} U_{ij} \sigma_{j} \rangle \neq 0$. 
As shown by Wegner \cite{wegner}, three  
(or 2+1) dimensional variants of
the ${\mathbb Z}_{2}/{\mathbb Z}_{2}$ model with couplings
($\beta, K)$ are equivalent to the same model at couplings
$(\beta^{*}, K^{*})$ related via the self-duality relations
\begin{eqnarray}
\exp(-2 \beta^{*}) = \tanh K, ~ \exp(-2K^{*}) = \tanh \beta.
\label{dual}
\end{eqnarray}

\section{Duality and The Lee Yang Theorem}
\label{LY}

As illustrated above, the matter coupled 
gauge theory can be re-interpreted
as a pure gauge theory with an
additional magnetic field applied.
Such an analogy immediately triggers
a certain intuition regarding 
the exclusion of phase transitions
in certain systems. In standard spin models 
governed by the classical action
\begin{eqnarray}
S = - \frac{1}{2} \sum_{ij} J_{ij} s_{i} s_{j} - \sum_{i} h s_{i},  
\label{LYE}
\end{eqnarray}
with $H$ the Hamiltonian no phase transition can occur when a symmetry 
breaking magnetic field ($h \neq 0$) 
is applied. It is clear that the 
local magnetization magnetization 
$\langle s \rangle \neq 0$ and
this goes hand in hand with 
an analytic free energy. 

Lee and Yang \cite{LY} proved that, in the thermodynamic 
limit, the partition
function cannot have zeros. This can 
be shown to imply an analytic free energy
for magnetic fields for which 
$|\mbox{Im} \{ h \}| < |\mbox{Re} \{ h \}|$
(with Im and Re the imaginary and real components
respectively). This may be extended to many
systems. Its generalization to a pure ${\mathbb Z}_{2}$ lattice
gauge action with a magnetic field applied on each 
gauge link (Eq.(\ref{z})) on a general hyper-cubic lattice of dimension $d$
has been done \cite{Dunlop}. However, as 
Eq.(\ref{z}) is equivalent to the general matter coupled gauge theory of
Eq.(\ref{intro}), this implies that the free energy $(- \ln Z)$
is analytic for all sufficiently large matter couplings $\beta$.
More precisely \cite{Dunlop}, 
if $\theta_{\Box} \equiv \tanh K$,
and 
\begin{eqnarray} 
\theta_{\mbox{link}} \equiv \tanh \Big[ \frac{ \mbox{Re} \{ \beta \} -
|\mbox{Im}\{
\beta\}|}{2 (d-1)} \Big],
\end{eqnarray} 
then the partition function
is non-vanishing and the free energy analytic
in the region 
\begin{eqnarray}
\theta_{\mbox{link}}^{4} \ge \theta_{\Box} + \theta_{\Box}^{-1} 
+ 3 -  \sqrt{(\theta_{\Box} + \theta_{\Box}^{-1}
+3)^{2} -1}.
\label{R1}
\end{eqnarray} 
Next we consider the (2+1) dimensional case
and then briefly remark on the 
(1+1) dimensional theory. For the three (or 2+1) dimensional case,
Eq.(\ref{R1})
explicitly reads
\begin{eqnarray}
\tanh^{4} \frac{\beta}{4} \ge \tanh K + \coth K + 3
\nonumber
\\ - \sqrt{\Big( \tanh K + \coth K +
3 \Big)^{2}-1}.
\label{dualD3}
\end{eqnarray}

Let us now insert 
the self duality relations Eqs.(\ref{dual})
to obtain 
\begin{eqnarray}
\tanh K = \exp[- 2 \beta^{*}],
\nonumber
\\
\tanh \frac{\beta}{4} = \Big(\frac{\sqrt{1+ \lambda} 
-  \sqrt{2\lambda}}{
\sqrt{1+ \lambda} +  \sqrt{2\lambda}}\Big)^{1/2},
\label{R2}
\end{eqnarray}
with
\begin{eqnarray}
\lambda \equiv \sqrt{1 - \exp[-4 K^{*}]}.
\end{eqnarray}
Inserting Eq.(\ref{R2}) into Eq.(\ref{dualD3})
gives a domain of analyticity of the free energy in
$(\beta^{*},K^{*})$.
The union of both domains is a region free of non-analyticities.
In particular, we find that for all plaquette couplings
\begin{eqnarray}
K < -\frac{1}{2} \ln \tanh [4 \tanh^{-1}(5 -\sqrt{24})^{1/4}] 
\label{Cdual}
\end{eqnarray}
with arbitrary matter coupling $\beta$, the partition function of the three
dimensional
${\mathbb Z}_{2}/{\mathbb Z}_{2}$ theory
has no zeros in the thermodynamic limit
and the free energy is analytic.
Along the $\beta=0$ line of Eq.(\ref{intro}) (the pure gauge only
theory), the value $K=K_{c}$ 
at which a confining transition occurs
may be inferred from the critical temperature
of the three dimensional Ising model. Within the confining
transition of the pure gauge theory (the action of 
Eq.(\ref{intro}) in the absence
of matter coupling- $\beta =0$),
the Wilson loop $W_{C} = \langle \prod_{ij \in C} U_{ij} \rangle$ 
for a large loop $C$, changes from an asymptotic
perimeter law behavior ($W_{c} \sim e^{-c_{1}l}$
with $l$ the perimeter of $C$ and $c_{1}$ a constant) for large plaquette 
couplings ($K>K_{c}$) to a much more rapidly decaying area law 
($W_{c} \sim e^{-c_{2} A}$ with $A$ the area
of the minimal surface bounded by $C$ and $c_{2}$ a constant)
for weak couplings $K<K_{c}$\cite{Kogut}. At $K=K_{c}$, the free
energy is non-analytic. By duality (Eqs.(\ref{dual})), 
the location of this non-analyticity in $K$ along the $\beta=0$
axis maps onto the location of non-analyticity associated with the 
transition within the 3D Ising model 
($S_{3D~Ising} = - \beta \sum_{\langle ij \rangle}
\sigma_{i} \sigma_{j}$) at its critical point $\beta= \beta_{c}$.
Following Eqs.(\ref{dual}), 
the relation between the 
two is  $\tanh \beta_{c}^{\mbox{3 D Ising model}}$ = $\exp(- 2
K_{c}^{\mbox{3 D Ising gauge}})$. Numerically, 
in the 3D Ising model $\beta_{c}^{\mbox{3 D
Ising model}} \simeq 0.22165$. This implies
that the critical value of $K$ within the pure ($\beta =0$ in 
Eq.(\ref{intro})) 3D Ising gauge theory 
is $K_{c}^{\mbox{3 D Ising gauge}} \simeq 0.761423$, e.g. \cite{ID}. 
The partition
function is non-zero and the free 
energy is analytic within the
region given by Eq.(\ref{Cdual})
which lies within the confining 
phase of the three dimensional 
${\mathbb Z}_{2}/{\mathbb Z}_{2}$ model for small $\beta$. 
Thus, as pointed out in a seminal
paper by Fradkin and Shenker
\cite{Fradkin}, the Higgs 
(large $\beta,K$) and the confining
phases (small $\beta,K$) are analytically 
connected. No phase transition 
need be encountered in going from 
one phase to the other.
Here we explicitly prove
this for the three dimensional
${\mathbb Z}_{2}/{\mathbb Z}_{2}$ model
with explicit rigorous numerical bounds
as in Eq.(\ref{Cdual}).

\bigskip

\begin{figure}[htb]
\vspace*{-0.5cm}
\includegraphics[angle=0,width=6.5cm]{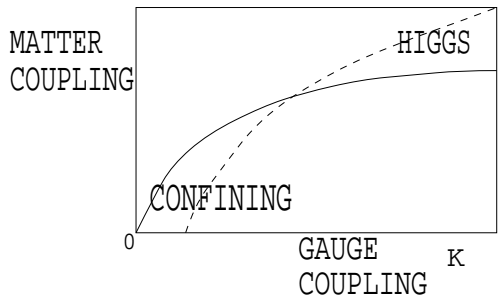}
\vspace*{-0.0cm}
\label{fig}
\caption{The region in the phase diagram
of the three dimensional ${\mathbb Z}_{2}/{\mathbb Z}_{2}$  
for which we prove that the partition
function is free of zeros and consequently
the free energy is analytic.
The horizontal axis 
is $K$- the strength
of the gauge field
and the vertical axis depicts
$\beta$ - the strength of the 
matter coupling. Both axis span the region
from 0 to $\infty$. The solid
line is the bound attained
from the Lee Yang theorem. The region 
above this curve is free
of non-analyticities. 
By duality, the region above 
dashed line is 
also free of non-analyticities.
Thus the union of both regions
is analytic. This connects the Higgs 
phase (high $\beta, K$)
to the confining phase
(low $\beta$ and $K$).
The intercept of 
dashed line with the 
$\beta =0$ axis is
found to be 
$K =  -\frac{1}{2} \ln \tanh [4 \tanh^{-1}(5 -\sqrt{24})^{1/4}]$,
which is within the confining
phase for small
$\beta$, as expected
(it cannot span the deconfining
phase as then a 
non-analyticity in the 
free energy or zero of
the partition function
would be encountered
at $K = K_{c}^{\mbox{3D Ising Gauge}}$).
}
\label{DUAL-FIG}
\end{figure}

The bound on a finite region of the phase diagram 
free of partition function zeros complements
the classic works of Marra and Miracle-Sole' \cite{Marra} 
that show that the small $\beta,K$ expansion 
of the free energy corresponding to 
Eq.(\ref{intro}) converges if $K$ is sufficiently
small irrespective of $\beta$, or if both $K$ 
and $\beta$ are sufficiently 
large. It is noteworthy that although
duality allowed us to generate stringent bounds,
the Lee-Yang theorem itself linked
points deep within the confining
phase $(\beta, K) \to (0,0)$ to
those in the Higgs phase $(\beta \gg 1, K \gg 1)$.
The extension of the Lee-Yang theorem to 
gauge theories other than ${\mathbb Z}_{2}/{\mathbb Z}_{2}$ 
is straightforward albeit technically more involved. 

\bigskip

\begin{figure}[htb]
\vspace*{-0.5cm}
\includegraphics[angle=0,width=6.5cm]{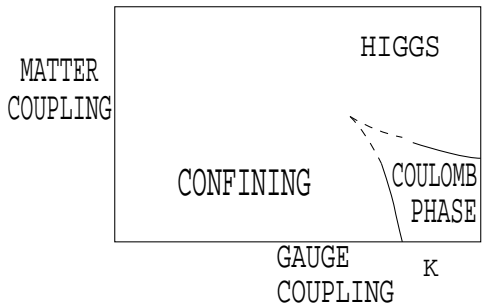}
\vspace*{-0.0cm}
\label{fig}
\caption{A schematic representation
of the phase diagram of the 
${\mathbb Z}_{2}/{\mathbb Z}_{2}$ 
theory in d=3 space dimensions. 
This phase diagram was proposed
by \cite{Fradkin}. The boundaries drawn in 
Fig.(\ref{DUAL-FIG}) are only 
bounds. The confining transition
extend deep beyond the line
implied by the Lee Yang 
theorem.}
\label{FS}
\end{figure}

We now examine the much more trivial
two dimensional incarnation
of the ${\mathbb Z}_{2}/{\mathbb Z}_{2}$
theory to illustrate
that it displays a single phase. By a duality mapping  
(see an explicit derivation in the Appendix), it is readily
seen that the partition function of the the two dimensional
${\mathbb Z}_{2}/{\mathbb Z}_{2}$ model at matter coupling
$\beta$ and gauge coupling $K$, is equal (up
to constants) to the 
partition function of the 
two dimensional Ising
model (of unit lattice spacing) 
given by Eq.(\ref{LYE}) of nearest neighbor exchange constant
\begin{eqnarray}
J_{ij} = \frac{1}{2} \ln \coth \beta ~ \delta_{|i-j|,1}
\label{J1.}
\end{eqnarray}
and uniform external magnetic field
\begin{eqnarray}
h = \frac{1}{2} \ln \coth K.
\label{h1.}
\end{eqnarray}

As the two dimensional Ising model
in a magnetic field
displays (via the Lee-Yang theorem)
no phase transitions, the  two dimensional 
${\mathbb Z}_{2}/{\mathbb Z}_{2}$
theory exhibits only a single phase
regardless of the strength of the 
couplings. Notwithstanding,
as we report towards the end of this Letter, 
the existence of a line of weak singularities
may be firmly established.

\section{General Considerations
for a single Higgs-Confining phase}

\begin{figure}[htb]
\vspace*{-0.5cm}
\includegraphics[angle=0,width=6.5cm]{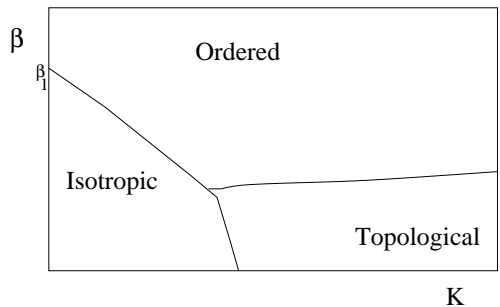}
\vspace*{-0.0cm}
\label{fig}
\caption{The phase diagram above was found by \cite{LRT}
for$ O(3)$ matter fields
coupled to ${\mathbb Z}_{2}$ gauge links in the context
of liquid crystals. Here, the confining, Higgs,
and Coulomb phases of the ${\mathbb Z}_{2}/{\mathbb Z}_{2}$ 
theory becomes
three different sharp phases (whose siblings
are respectively denoted
in the above as ``Isotropic'',  ``Ordered'',
and ``Topological''.) We prove, by employing self-duality
of the ${\mathbb Z}_{2}/{\mathbb Z}_{2}$ theory that 
a phase digram having three phases
such as that of $O(3)$ matter
coupled to ${\mathbb Z}_{2}$ gauges
shown above is impossible. 
In the ${\mathbb Z}_{2}/{\mathbb Z}_{2}$ theory, phase boundaries
may only terminate on
the $\beta=0$ or $K=\infty$ axis-
no phase boundary can separate the Higgs
and confining phases.}
\label{impossible}
\end{figure}

Next, we avoid the use rigorous Lee-Yang bounds
and ask ourselves what statements can
be made regarding the phase diagram
on general principle alone both in the
presence and absence of dualities.
First, we illustrate that a phase diagram
such that shown in Fig.(\ref{impossible})
is impossible for the ${\mathbb Z}_{2}/{\mathbb Z}_{2}$ theory. 
The phase diagram depicted in 
Fig.(\ref{impossible}) was proposed for the 
very different theory of $O(3)$ matter fields coupled to 
${\mathbb Z}_{2}$ 
by \cite{LRT} in their beautiful theory
of liquid crystals. 

The proof of the impossibility of such a 
phase diagram for a ${\mathbb Z}_{2}/{\mathbb Z}_{2}$ theory 
and the necessity
of having a single Higgs-confinement phase is quite straightforward. 
As the ${\mathbb Z}_{2}/{\mathbb Z}_{2}$ theory is self-dual 
(see Eqs.(\ref{dual})), the phase
diagram must look the
same under the duality
transformation. The phase
boundaries where the partition
function vanishes, $Z(\beta,K)=0$,
must be the same as those
where $Z(\beta^{*},K^{*})=0$.
A phase diagram such as 
Fig.(\ref{impossible}) does not
satisfy self-duality. A critical
line emanating from $(\beta= \beta_{1},K=0)$ immediately implies
a line of singularities
emanating from $(\beta^{*}= \infty,K^{*}= - \frac{1}{2} \ln \tanh \beta_{1})$.
If $Z(\beta^{*},K^{*}) =0$ along this line then, as
the functional form for $Z(\beta^{*},K^{*})$
is equivalent to that of $Z(\beta,K)$ with merely
the coupling constant tuned to different values, 
$Z(\beta,K)$ must also have a line of zeros
emanating from $(\beta= \infty, K = - \frac{1}{2} \ln \tanh \beta_{1})$
and the phase diagram must possess, at least, another
line of singularities. The same would apply to a line of singularities
starting from $(\beta= \infty, K = K_{1})$
which is easily excluded. 

Next, we look at the physics of the models
in their limiting incarnations. At $K=0$ the partition function
of the ${\mathbb Z}_{2}/{\mathbb Z}_{2}$
theory is trivially $Z=(2\cosh \beta)^{Nd}$ 
with $N$ the number of lattice sites. Here, the
system is simply that of free 
bonds in a magnetic field and no singularities
can occur at any value of $\beta=\beta_{1}$ with $K=0$.
Self-duality then implies that no singularities can occur in the self-dual
${\mathbb Z}_{2}/{\mathbb Z}_{2}$ theory at $\beta= \infty$ and any finite 
value of $K$.

Putting all of the pieces together, by employing self-duality,
and the absence of 
singularities at
$K=0$, within the ${\mathbb Z}_{2}/{\mathbb Z}_{2}$ theory,
lines of singularities in the phase
diagram can 
only originate from $(\beta=0,K=K_{c})$
or from $(\beta=\beta_{c},K=0)$ 
(with possibly more than
one value of $K_{c}$ ($\{K_{c,i}\}$) and/or $\beta_{c}$) 
or form closed loops or lines of
transitions terminating in the bulk. 
States with $\beta =0$
and $K < \min_{i}\{K_{c,i}\}$
and those with $K=\infty$ and $\beta < \max_{j} \{\beta_{c,j}\}$
must be analytically connected
to each other. In the 
standard spin ($K=0, \beta= \beta_{c}$)
and gauge models
$(\beta=0, K = K_{c})$ 
only a single 
critical value appears.
The Higgs and confining phases must, asymptotically,
be one and the same. A singularity
anywhere along the line 
$K=0^{+}$ is excluded
in the self-dual theory
as that limit corresponds
to $\beta \to \infty$
which is completely
ordered ($z_{ij} =1$)
and no transitions occur.
This proves the celebrated result
of \cite{Fradkin}.

We now examine the situation in general non self-dual
theories in which the matter
fields ($\sigma_{i}$) are in a subgroup of
the gauge group (the group $G$ such that all links $U_{ij} \in G$). 
(This situation does not encompass theories such as
those described by \cite{LRT}, \cite{SF}.) 
In such instances, the bond variables $z_{ij} = \sigma_{i}^{*} 
U_{ij} \sigma_{j}$ 
are elements of $G$. Similar to Eq.(\ref{z}),
we may parameterize the action in terms of 
the gauge invariant 
link variables $\{z_{ij}\}$.
In what follows, we focus for concreteness on 
$U(1)$ (or $O(n=2)$) theories. 
First, we note that along the $K=0$ axis, the pure non-interacting
links in the effective
magnetic field $\beta$
(leading to $Z=  (I_{n/2-1}(\beta))^{Nd}$ [with $I_{n/2-1}$ 
a Bessel function
of order $(n/2-1)$]
for $O(n)$ fields) display
no singularities in the
free energy.
Along the $\beta = \infty$
axis, irrespective of
the value of $K$,
all the gauge invariant
bonds $z_{ij}$ 
in the $U(1)$ theory are pinned to 1.
No transitions occur as $K$
is varied along the $\beta=\infty$
line as all bond variables
are already frozen at their maximally magnetized 
unit value. In fact, increasing $K$ for
finite $\beta$ can only
make this magnetization
stronger. The partition function has no
dependence on $K$ along this
line. Thus, we see that in general
no phase boundaries 
can traverse the 
$\beta = \infty$
or the $K=0$ line
even in the absence
of self-duality
and Lee-Yang results
which allow us to 
make matters more
elegant and provide
rigorous numerical bounds.
Thus, the Higgs and confining phases
are one and the same for
all of these theories.
We must nevertheless mention 
that in non self dual
theories, relying only
on the above we cannot
immediately exclude a transition 
boundary ending in the 
bulk at $K=0^{+}$. To exclude
this for different individual
theories, we need to examine
the radius of convergence
in $K$.

\section{Establishing new percolation crossovers by duality} 
\label{percolationsection}

\begin{figure}[htb]
\vspace*{-0.5cm}
\includegraphics[angle=0,width=5cm]{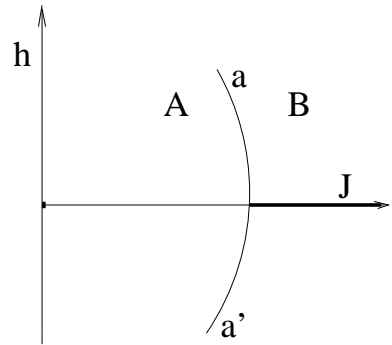}
\vspace*{-1.0cm}
\label{fig}
\bigskip
\bigskip
\caption{A caricature of the original \cite{kertesz}
phase diagram proposed by kertesz and later
verified in detail 
by \cite{adler}.  
The $J$ and $h$ axes parameterize
the classical two dimensional Ising 
Hamiltonian $S = - J \sum_{\langle ij \rangle} 
s_{i} s_{j} - h  \sum_{i} s_{i}$.
The low temperature solid line along
the $h=0$ axis denotes the usual 
first order transition, while the fainter
lines denote Kertesz transitions.
Here, in phase A there is an exponentially
rare number of droplets whose spin points opposite
to the applied field $h$ with a finite surface tension.
In Phase B, the surface tension of oppositely oriented 
spin droplets vanishes. The lines $a$ and $a'$ denote
the droplet cluster transition across which the 
surface tension ($\Gamma$ of Eq.(\ref{kerteq}))
drops to zero.}
\label{KERTESZ}
\end{figure}

\begin{figure}[htb]
\vspace*{-0.5cm}
\includegraphics[angle=0,width=4cm]{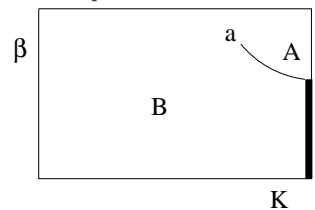}
\vspace*{-1.0cm}
\label{fig}
\bigskip
\bigskip
\caption{The phase diagram of the two dimensional matter coupled
Ising gauge theory (Eq.(\ref{intro}) in $d=2$ dimensions)  
as derived from Fig.(\ref{KERTESZ}) following the 
duality transformations of Eqs.(\ref{J1.},\ref{h1.}).
Albeit appearances, this gauge theory has a percolation 
transition. As we establish
here, the two dimensional
Ising theory of Eq.(\ref{intro}), by an application of the duality
relations of Eqs.(\ref{J1.}, \ref{h1.}),
a percolation transition separates phase $B$
(spanning the confining regime) and phase $A$
(which overlaps with the Higgs phase).
The solid line along the $K = \infty$ axis
signifies the disordered phase
of the simple ferromagnet.
(At $K= \infty$, Eq.(\ref{intro})
reduces to $S = - \beta \sum_{\langle ij \rangle} \sigma_{i} \sigma_{j}$,
the action of a ferromagnet.)}
 \label{LATTICE}
\end{figure}

With all stated thus far, it would appear
that the single Higgs-confining phase is 
one bulk phase and no transitions occur
within it. We now illustrate that 
this is not the case-
at least not within the 
simplest of all matter coupled
lattice gauge theories-
the two dimensional ${\mathbb Z}_{2}/{\mathbb Z}_{2}$ theory
which we now show to possess a richer phase diagram than
anticipated (a single phase). With no matter, as
is well known e.g. \cite{Kogut}, \cite{polyakov}, 
the pure two dimensional ${\mathbb Z}_{2}$ gauge
theory given by the plaquette term of 
Eq.(\ref{intro}) is equivalent (by a trivial gauge fix, e.g. $U_{ij} =1$ 
on all horizontal bonds in the plane) to a stack of decoupled
one dimensional Ising chains (all of which are horizontal Ising 
chains formed by the vertical bonds in the gauge alluded to here).
As Ising chains are disordered at any finite coupling,
the two dimensional ${\mathbb Z}_{2}$ gauge
theory is always confining. 

Now let us introduce matter coupling (a finite $\beta$
in Eq.(\ref{intro})) and 
consider the following thought experiment: we
color every appearance of $z_{ij} = \pm 1$ in
the two dimensional ${\mathbb Z}_{2}/{\mathbb Z}_{2}$ theory 
by one of two colors and ask ourselves whether
the bonds of a uniform sign (the $z_{ij} = 1$ bonds
for $\beta >0$) percolate, upon a trivial mapping, across the sample and if so
whether a transition between a percolative and non-percolative 
clusters can exist within the single Higgs-confining phase. 
Although this question is very general, we can 
make easy progress and establish
rigorous results by relying
on the exact duality of Eqs.(\ref{J1.},\ref{h1.})
to the well studied model of
a two dimensional Ising magnet in a magnetic field.
Some time ago, Kertesz argued 
\cite{kertesz} that although
there might not be (via the Lee-Yang theorem)
any thermodynamic singularities in various 
spin models when subjected 
to an external magnetic 
field, sharp crossovers related to  
vanishing surface tension (of droplets
of oppositely oriented spins) and a change of character of 
the high field series (for quantities such as the free energy
or magnetization) occur. 
As is well established by now, this crossover 
may be discerned by the percolation of clusters  
constructed via the Fortuin-Kastelyn representation
\cite{FK}. We would like to suggest 
that the massive character of
the photons in the confining/Higgs
regimes may reflect such 
a difference. Here,
the spins 
of \cite{kertesz} are replaced
by a functional of the gauge invariant meson variables 
$z_{ij} \equiv \sigma_{i}^{*} U_{ij} \sigma_{j}$. 
For the 1+1 dimensional ${\mathbb Z}_{2}/{\mathbb Z}_{2}$ theory,
we now readily establish this result: 
By Eqs.(\ref{J1.},\ref{h1.}), the 
d=1+1 dimensional ${\mathbb Z}_{2}/{\mathbb Z}_{2}$ theory
may be directly mapped to a two dimensional
Ising model in a magnetic field given by Eq.(\ref{LYE}).
However, as established by \cite{adler} 
the two dimensional Ising model in a magnetic
field displays a Kertesz line. 
In general dimension $d$,
with $n_{V}$ the number of drops of volume $V$ whose
spins are oppositely oriented
(those opposing the external magnetic field $h$
in Eq.(\ref{LYE})), we have, for large $V$,
the leading order relation
\begin{eqnarray}
\ln  n_{V} \sim - 2h V - \Gamma V^{(d-1)/d}. 
\label{kerteq}
\end{eqnarray}
Here, the surface tension $\Gamma$ vanishes in one 
phase (phase B of Fig.(\ref{KERTESZ}))
while it is finite in the other (phase
A in Fig.(\ref{KERTESZ})) \cite{kertesz}.
Equivalently, this crossover may be ascertained 
via the examination of
the radii of expansion \cite{adler}
in $\mu \equiv e^{-2h}$ (see Eq.\ref{LYE}) for the magnetization
\begin{eqnarray}
\langle s \rangle = 1 - 2 \sum_{V} V L_{V}(u) \mu^{V},
\end{eqnarray}
where $u \equiv e^{-2J}$, and $L_{V}$ is a polynomial in $u$. 
Although for any finite $h$, the radius of convergence
in $\mu$ is finite (as indeed no transitions occur
by the Lee-Yang theorem),
the radius of convergence increases across the percolation line
(appearing as jumps in \cite{adler}).
For couplings $J$ larger than the percolation
threshold $J>J_{p}$, the radius of convergence
in $\mu$ is up to $\mu =1$ (i.e. it is convergent for 
all $h \ge 0$ in Eq.(\ref{LYE})) 
and to a larger value $\mu >1$ for $J<J_{p}$ (the surface tension 
free regime)- up to finite negative values of $h$ \cite{adler}. 
Upon dualizing (Eqs.(\ref{J1.}, \ref{h1.})), this implies an identical
crossover in the single plaquette expectation value of 
$\langle z_{ij}z_{jk}z_{kl}z_{li} \rangle$
(which is none other than the minimal Wilson
loop $\langle U_{ij} U_{jk} U_{kl} U_{li} \rangle$) 
when expanded in powers
of $\tilde{\mu} \equiv \tanh K$
for fixed $\tilde{u} \equiv \tanh \beta$.
The transition is discerned by 
the convergence of the single plaquette
expectation value up to negative $K$ values.

Taken together, the duality relations of Eqs.(\ref{J1.},\ref{h1.}) and the
firm results of \cite{adler} prove, for the first time,  
that the two dimensional ${\mathbb Z}_{2}/{\mathbb Z}_{2}$
theory must also exhibit a Kertesz line. A sketch of
the original phase diagram
of Kertesz \cite{kertesz} and
its new gauge theory dual are depicted
in Figs.(\ref{KERTESZ}, \ref{LATTICE}).
Percolation transitions established
here for the two dimensional ${\mathbb Z}_{2}/{\mathbb Z}_{2}$
theory and others speculated elsewhere
might be linked to infinite Wilson loop
like observables \cite{explain, wilson}.

\section{Conclusions}

In conclusion, we illustrate that a phase diagram
of a gauge theory coupled to matter can be proved 
directly and stringent numerical bounds provided. Our methods 
reaffirm the seminal result of Fradkin and Shenker 
\cite{Fradkin}. We further remarked on extensions
of this result. Our results suggest that
the existence of a single Higgs-confining phase in some 
theories (as mandated via a
generalized Lee-Yang theorem in the ${\mathbb Z}_{2}/{\mathbb Z}_{2}$
theory and strongly hinted by general considerations
in other general instances) can often be viewed as the analogue of
the absence of phase transitions in
spin systems subjected to an external
magnetic field. Similar to such 
spin systems, we speculate that 
the locus of gauge and matter couplings $(K, \beta)$ 
at which a correlated percolation of clusters (given by 
an effective spin state related to gauge invariant bonds variables
$z_{ij} \equiv  \sigma_{i}^{*} U_{ij} \sigma_{j}$) occurs 
may constitute an analogue of the Kertesz line
known in such spin systems \cite{kertesz}. 
We establish the validity
of this anticipation and {\bf{ the existence
of a Kertesz line}} within a simple gauge 
theory harboring a single confining
phase- the 1+1 dimensional  
${\mathbb Z}_{2}/{\mathbb Z}_{2}$ theory.
Possible manifestations of this effect
for more physically pertinent 
higher group gauge fields 
in $d=4$ remain a speculation. 

\section{Acknowledgments}
I am indebted to Jan Zaanen and Asle Sudb\o~ 
for many conversations and for another work in progress \cite{NZS}. 
This work was partially supported by 
US DOE under LDRD X1WX and by FOM.

\appendix

\section{Derivation of the duality of the ${\mathbb Z}_{2}/{\mathbb Z}_{2}$ 
theory}
\label{app1}

The ${\mathbb Z}_{2}/{\mathbb Z}_{2}$ theory of 
Eq.(\ref{z}) in $d=2$ dimensions
is dual (via Eqs.(\ref{J1.},\ref{h1.})) to the two dimensional
Ising in an external
magnetic field of Eq.(\ref{LYE}). This duality allowed us
to prove the existence of a sharp percolation crossover
(a Kertesz \cite{kertesz} line) within the 
Higgs-confining phase of the simplest of all matter
coupled gauge theories- the 
two dimensional ${\mathbb Z}_{2}/{\mathbb Z}_{2}$ theory.
The existence of a duality between the 
two dimensional Ising model in a magnetic field and the
two dimensional ${\mathbb Z}_{2}/{\mathbb Z}_{2}$ theory 
was noted in \cite{Fradkin}. As this 
duality is pivotal in proving our new percolation crossovers 
(section(\ref{percolationsection})) even in this simplest
of all matter coupled gauge theories, we explicitly 
illustrate its derivation below. 

\begin{figure}
\includegraphics[width=5.2cm]{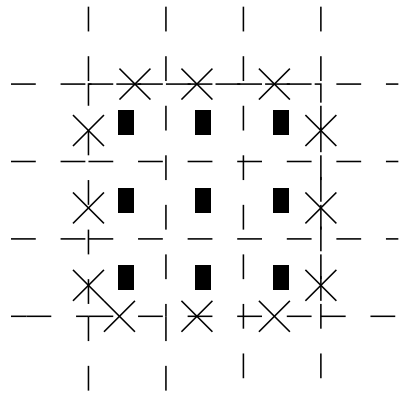}
\caption{
A contribution to the
low coupling series of
the two dimensional ${\mathbb Z}_{2}/{\mathbb Z}_{2}$ action.
The centers of plaquettes
are labeled by the solid rectangles.
The crosses (x) denote energetic bonds (gauge invariant 
bond variables $z$ of text) residing on
the perimeter of the contour.
}
\label{DUAL1-FIG}
\end{figure}

In what follows, we employ series expansions (a standard approach
for deriving many dualities) in the high and low coupling limits
to show that the high and low coupling regimes of the two disparate
models (the two dimensional ${\mathbb Z}_{2}/{\mathbb Z}_{2}$ theory
of Eq.(\ref{z}) and the two dimensional Ising model in a magnetic field
of Eq.(\ref{LYE})) become
one and the same upon a change of variables (the duality
transformation). Hereafter, we set in Eq.(\ref{LYE})
the exchange constant $J_{ij} = J \delta_{|i-j|,1}$ .
We start by expanding the partition function
\begin{eqnarray}
Z = \sum_{ \{z_{ij}\} } \exp[-S],
\label{partitionf}
\end{eqnarray} in a small $\beta,K$ 
(``high temperature'')
series. In Eq.(\ref{partitionf}), the action
$S$ is given by Eq.(\ref{z}) and the summation in 
Eq.(\ref{partitionf}) spans
all gauge invariant bond variables 
($z_{ij} = \pm 1$ on all nearest neighbor links $\langle i j \rangle$).  
To attain the low coupling expansion of the partition function
$Z$ of Eq.(\ref{partitionf}), we employ the identities
\begin{eqnarray}
\exp[\beta z] = \cosh \beta(1+ z \tanh \beta), \nonumber
\\ \exp[K zzzz] = \cosh K( 1 + zzzz \tanh K), 
\end{eqnarray}
to obtain a polynomial expansion in 
\begin{eqnarray}
x = \tanh K, ~ y = \tanh \beta.
\end{eqnarray}
With these elements in tow, the partition
function of Eq.(\ref{partitionf}) becomes a
sum of diagrams. These diagrams (Fig.(\ref{DUAL1-FIG}))
correspond to drawing closed
contours in the plane
and counting the number
of dual lattice sites (the centers of plaquettes
surrounded by 4 gauge links- corresponding to the 
plaquette terms $zzzz$ stemming from the exponentiation 
of the second term in the action of Eq.(\ref{z})- 
which are labeled by the solid rectangles)
and the number of bonds ($z$), obtained from exponentiation
of the first term of Eq.(\ref{z}),
labeled by the crosses
residing on the contour boundaries. The sum over
all values of $z_{ij} = \pm 1$ allows only diagrams containing
closed loops in which each bond ($z_{ij}$) appears to 
an even powers (all other diagrams necessarily have
at least one bond which appears an odd number of times 
and therefore leads to zero once the sum over $z_{ij}= \pm 1$ is 
performed). The sum over each bond, $\sum_{z_{ij} = \pm 1} z_{ij}^{n_{ij}} = 2$
for all even $n_{ij}$ ($n_{ij} = 0, 2$). 
All in all, the series for the partition
function becomes
\begin{eqnarray}
Z =  4^{N} (\cosh K)^{N}  (\cosh \beta)^{2N} \sum_{\mbox{closed loops}}  
x^{A} y^{|C|},
\end{eqnarray}
where $A$ denotes the net area enclosed
by the set of loops $C$ and, $|C|$ marks the
net perimeter of all closed
loops making up the joint contour
$C$.

\begin{figure}
\includegraphics[width=5.2cm]{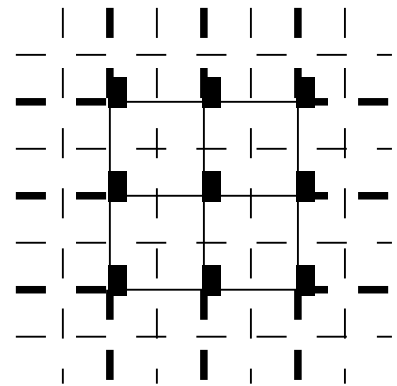}
\caption{The corresponding contribution to
the strong coupling (``low temperature'') series
of the two dimensional
Ising model in a
magnetic field. The flipped spins are
marked by black rectangles
with the 
bad energetic bonds that
the flipped spins generate
along their perimeter 
marked by a thick dashed line.
The bonds of the dual lattice are marked 
by a thin dashed
line.}
\label{DUAL2-FIG}
\end{figure}
If we expand the partition 
function corresponding to 
the action of 
Eq.(\ref{LYE}) about $J \to \infty$
(corresponding to flipping the spins
$s_{i}$ from their infinite coupling (``zero temperature'') 
ground state value of one ($h>0$))
then we will obtain a polynomial 
expansion in 
\begin{eqnarray}
\tilde{x} = \exp[-2 h],~ \tilde{y} = \exp[-2J].
\end{eqnarray}
Explicitly, the partition function
reads 
\begin{eqnarray}
\tilde{Z} = \tilde{Z}_{0}[\sum_{\mbox{clusters of flipped spins}} 
(\tilde{x})^{A}
(\tilde{y})^{|C|}],
\end{eqnarray}
where $\tilde{Z}_{0}$ is the zero temperature (infinite $h$ and $J$)
partition function, $A$ is the net area of all clusters
of flipped spins $(s_{i} = -1$), and $|C|$ is
the perimeter of all clusters of flipped
spins. In Fig.(\ref{DUAL2-FIG}), 
a simple cluster of flipped spins
is shown. The flipped spins are
marked by black rectangles
(each flipped spin incurs
a Boltzmann energy penalty
of $\tilde{x}$), and the 
bad energetic bonds that
the flipped spins generate
along their perimeter (perpendicular to the domain wall)
are marked by a thick dashed line. 
In Fig.(\ref{DUAL2-FIG}),
the bonds of the dual lattice are marked 
by a thin dashed
line. Note the obvious
one to one relation between
the cluster of flipped
spins of Fig.(\ref{DUAL2-FIG})
to the high temperature limit
in Fig.(\ref{DUAL1-FIG}). 
The polynomials for 
the partition functions
$Z$ and $\tilde{Z}$ 
in ($x,y$)
and $(\tilde{x},\tilde{y})$
respectively are identical.
The partition function of the the two dimensional
${\mathbb Z}_{2}/{\mathbb Z}_{2}$ model at matter coupling
$\beta$ and gauge coupling $K$, is equal (up
to constants) to the 
partition function of the 
two dimensional Ising
model of exchange constant
and magnetic field given by Eqs.(\ref{J1.},\ref{h1.}).
As the two dimensional Ising model in a magnetic
field of Eq.(\ref{LYE}) exhibits no phase transition
and thus no non-analyticities for any non-zero $h$
(corresponding to any finite $K$ in the action of Eq.(\ref{z})
on the square lattice), the radii of expansion of the series derived 
above are infinite and the duality transformations 
of Eqs.(\ref{J1.},\ref{h1.}) hold throughout.


\begin{thebibliography}{10}
\bibitem{Josephson} These meson like variables $(\{z_{ij}\}$)
have clear physical meaning. For instance, within 
higher group incarnations of the 
action of Eq.(\ref{z}) wherein $z_{ij} = \sigma_{i}^{*} U_{ij}
\sigma_{j}$, e.g. the theory of $U(1)$ matter
coupled to $U(1)$ gauge (the $U(1)/U(1)$ theory),
$z_{ij}$ is directly linked to the gauge
invariant current between lattice sites $i$ and $j$.
Within the $U(1)/U(1)$ theory describing a hypercubic array 
of coupled superconducting grains (whose superconducting order 
parameter is given by a phase ($\sigma_{i} = e^{i \theta_{i}}$))
in an external magnetic field (governed by the $U(1)$ gauge 
field $\vec{A}$ or, on the hypercubic lattice with 
$U_{ij} = \exp[i \int_{i}^{j} 
\vec{A} \cdot d \vec{r}]$), these gauge invariant fields 
$z_{ij} = \exp[i \phi_{ij}]$, with the Josephson 
current between sites $i$ and $j$ given
by $J_{ij} = \sin \phi_{ij}$. By considering
the general relation giving the current
as the variational derivative of the action
with respect to the gauge potential, $J = \frac{\delta S}{\delta A}$,
it is readily seen that this identification
is far more general.
\bibitem{explain} When the matter fields carry the fundamental unit of charge
(which is not the situation in certain of the theories
of \cite{wilson}) ,
the Wilson loop $W_{C} = \langle \prod_{ij \in C} U_{ij} \rangle$
along any contour $C$ (including the boundaries of the system) 
can be expressed in terms of the gauge invariant fields defined here
as $W_{C} = \langle \prod_{ij \in C} z_{ij} \rangle$.
\end{thebibliography}
\end{document}